\documentclass[prl,twocolumn,aps]{revtex4}
\usepackage{graphicx,amssymb,amsbsy,bm,amsmath,psfrag,dsfont}
\usepackage{slashed}
\usepackage{hyperref} \usepackage[dvips]{color}

\begin{document}

\newcommand{\be}{\begin{equation}}
\newcommand{\ee}{\end{equation}}
\newcommand{\bea}{\begin{eqnarray}}
\newcommand{\eea}{\end{eqnarray}}
\newcommand{\no}{\noindent}

\newcommand{\la}{\lambda}
\newcommand{\si}{\sigma}
\newcommand{\vk}{\vec{k}}
\newcommand{\vx}{\vec{x}}
\newcommand{\om}{\omega}
\newcommand{\Om}{\Omega}
\newcommand{\ga}{\gamma}
\newcommand{\Ga}{\Gamma}
\newcommand{\gaa}{\Gamma_a}
\newcommand{\al}{\alpha}
\newcommand{\ep}{\epsilon}
\newcommand{\app}{\approx}
\newcommand{\uvk}{\widehat{\bf{k}}}
\newcommand{\OM}{\overline{M}}

\title{Leptogenesis and CPT Violation}
\author{Chiu Man Ho} \email{chiuman.ho@vanderbilt.edu}
  \affiliation{Department of
  Physics and Astronomy, Vanderbilt University, Nashville, TN 37235, USA}

\date{\today}

\begin{abstract}
We construct a model in which neutrinos and anti-neutrinos
acquire the same mass but slightly different energy dispersion relations.
Despite CPT violation, spin-statistics is preserved. We find that leptogenesis can be easily
explained within this model, without upsetting the solar, atmospheric and reactor neutrino data.
Leptogenesis occurs without lepton number violation and the non-equilibrium condition. We consider only three active
Dirac neutrinos, and no new particles or symmetries are introduced.
\end{abstract}

\maketitle

\emph{Introduction.} \,\,
Despite the crucial significance of CPT symmetry in the conventional quantum field theory,
it has been shown that string interactions may induce couplings between Lorentz tensors and fermions in the low-energy 4D
effective lagrangian \cite{Alan}. When the appropriate components of these Lorentz tensors acquire non-zero vacuum expectation values,
they lead to a spontaneous CPT violation.

Recently, the result from MINOS suggests a tension between the oscillation parameters for $\nu_\mu$ and $\bar{\nu}_{\mu}$
disappearance \cite{Vahle}. More precisely, at 90\% confidence level, it reports that:
$|\,\Delta m_{32}^2\,| = 2.35^{+0.11}_{-0.08} \times 10^{-3} \,\textrm{eV}^2$\,  and \,
$|\,\Delta \overline{m}_{32}^2\,| = 3.36^{+0.45}_{-0.40} \times 10^{-3} \,\textrm{eV}^2$ \,,
together with $\sin^2 ( \,2 \theta_{23}\,) > 0.91$ and $\sin^2 ( \,2 \overline{\theta}_{23}\,) = 0.86 \pm 0.11$. This
substantial difference in the neutrino and anti-neutrino mass-squared splittings, if persists, may indicate CPT
violation in the neutrino sector.

The quest to explain MINOS could serve as a phenomenological motivation for CPT violation
in the neutrino sector. In fact, there have been some works along this direction \cite{CPTV}.
But this motivation is not unique, as addition of sterile neutrinos \cite{Nelson} or non-standard neutrino interactions \cite{Joachim}
may also provide an explanation.

An earlier suggestion for CPT-violating neutrinos was due to the unresolved neutrino data in the Liquid Scintillator Neutrino
Detector (LSND) \cite{LSND}. In LSND,  the flux of $\bar{\nu}_\mu$ lies in the energy range $ 20 \,\textrm{MeV} < E < 52.8 \,\textrm{MeV}$,
with an average about 40 MeV. The transition probability of $\bar{\nu}_{\mu} \rightarrow \bar{\nu}_{e}$ has been measured.
The data indicate a 3.8 $\sigma$ excess of $\bar{\nu}_e$ events, which, if interpreted as originating from
$\bar{\nu}_{\mu} \rightarrow \bar{\nu}_e$ oscillation, would imply a mass-squared splitting larger than 0.1 eV$^2$. This is at odd with
the data from SNO \cite{SNO}, KamLAND \cite{KamLAND} and Super-Kamiokande \cite{SuperK}. It was then suggested that CPT-violating neutrinos
may reconcile all the data \cite{LSNDCPTV}.

In MiniBooNE, the energy of neutrinos or anti-neutrino is of order GeV \cite{MiniNull}.
Recently, the $\nu_\mu \rightarrow \nu_e$ search in MiniBooNE has found no evidence for an excess of $\nu_e$ in the
$\nu_{\mu} \rightarrow \nu_{e}$ search \cite{MiniNull}, and it reveals that a mass-squared splitting smaller than
$0.1 \,\textrm{eV}^2$ is allowed. However, their $\bar{\nu}_\mu \rightarrow \bar{\nu}_e$ study appears to be consistent
with LSND \cite{Mini}, and a mass-squared splitting of at least $10^{-2} \,\textrm{eV}^2$ is required.  Again, this seems to
suggest CPT violation in the neutrino sector.

In this article, we would like to provide a new and simple model with only three CPT-violating active Dirac neutrinos, which
explains leptogenesis without lepton-number violation. At the same time, we show that this model is consistent with all of solar,
atmospheric and reactor neutrino data.
\\


\emph{Neutrinos and CPT Violation.} \,
Parallel to the most popular framework advocated in \cite{Alan_nuCPTV}, we propose a new type of Lorentz and CPT violations in the
neutrino sector. Our model is described by the following Lagrangian:
\bea
\label{L_flavor_covaraint}
L &=& \bar{\nu}_{\alpha} \left(\,i\,\delta_{\alpha\beta}\,\slashed{\partial}- m_{\alpha \beta}\,\right) \nu_{\beta}
+\,\bar{\nu}_{\alpha}   \,\lambda_{\alpha\beta} \,\gamma^0 \,T^{AB}\,\hat{\partial}_A\,\partial_B\,
 \nu_{\beta}\,, \nonumber \\
\eea
where $\alpha,\beta =e, \mu, \tau$,\, $m_{\alpha \beta}$ is the mass mixing matrix, $\lambda_{\alpha \beta}$
are dimensionless parameters characterizing kinetic mixings
between different flavors of neutrinos, $A,B=0,1,2,3$ are spacetime indices and $T^{AB}$ is a constant background
tensor which breaks Lorentz invariance. For a given scalar function $f$, we define the unit operator $\hat{\partial}_A$
as the following:
$\hat{\partial}_A\,f =\left(\,\partial_0\,f,~ \hat{\nabla}\, f\,\right) =\left(\,\partial_0\,f,~\frac{\vec{\nabla}\,f}{|\,\vec{\nabla}\,f\,|}\,\right)\,.$
Under Lorentz transformation, the unit operator transforms as $\hat{\partial}_{A'} =\frac{\partial\, x^A}{\partial\, x^{A'}}\,\hat{\partial}_A$.
The components of $T^{AB}$ can be chosen such that $T^{AB}=\delta^{AB}$ if $A,B=1,2,3$
and $T^{AB} =0$ otherwise. In this way, \eqref{L_flavor_covaraint} reduces to
\bea
\label{L_flavor}
L &=& \bar{\nu}_{\alpha} \left(\,i\,\delta_{\alpha\beta}\,\slashed{\partial}- m_{\alpha \beta}\,\right) \nu_{\beta}
+\,\bar{\nu}_{\alpha}   \,\lambda_{\alpha\beta} \,\gamma^0 \, \hat{\nabla} \cdot \vec{\nabla} \, \nu_{\beta} \,.
\nonumber \\
\eea
As the neutrino field will be expanded as a linear combination of the plane-wave solutions $e^{\pm \,i\, \vec{p}\cdot \vec{x}}$,
the unit gradient operator $\hat{\nabla}$ essentially operates as
\bea
\hat{\nabla}\; e^{\pm \,i\, \vec{p}\cdot \vec{x}} \equiv
\frac{\vec{\nabla}\; e^{\pm \,i\, \vec{p}\cdot \vec{x}} }{|\,\vec{\nabla}\; e^{\pm \,i\, \vec{p}\cdot \vec{x}}\,|}
=\pm\,i\, \hat{p} \; e^{\pm \,i\, \vec{p}\cdot \vec{x}}.
\eea
Obviously, the operator $\hat{\nabla}$ is ill-defined at zero-momentum $\vec{p}=\vec{0}$. However, the
composite operator
$\hat{\nabla} \cdot \vec{\nabla}$ is well-defined because in the
limit $ p \rightarrow 0$, $\hat{\nabla} \cdot \vec{\nabla}\,e^{\pm \,i\, \vec{p}\cdot \vec{x}}$ vanishes.
This is true regardless of the ordering of \,$\hat{\nabla}$\, and \,$\vec{\nabla}$,\, namely\,
$\vec{\nabla} \cdot \hat{\nabla}\,e^{\pm \,i\, \vec{p}\cdot \vec{x}}$ also vanishes
in the limit $p \rightarrow 0$.

We emphasize that the Lagrangian \eqref{L_flavor} is hermitian and renormalizable. The new operator with $\lambda_{\alpha\beta}\,\gamma^0$
breaks C but preserves P and T, and so violates CPT. Since CPT violation implies Lorentz violation \cite{Greenberg},
this operator also breaks Lorentz invariance. While the particle Lorentz invariance is broken, the observer Lorentz invariance is
preserved and so this new operator is consistent with the analysis provided by \cite{Lehnert}. For a symmetric $\lambda_{\alpha\beta}$, this
operator will
be identically zero if we consider Majorana neutrinos, because they do not have a vector current.
If $\lambda_{\alpha\beta}$ is anti-symmetric, Majorana neutrinos will be allowed but CPT will no longer be violated.
Since we are interested in neutrino CPT violation, we are essentially considering Dirac neutrinos in this article.

We assume that the mass mixing and kinetic mixing matrices commute with each other, and so we can diagonalize them simultaneously.
Upon diagonalization by the usual unitary transformation, we obtain neutrino mass eigenstates and the Lagrangian becomes
\bea
\label{L_mass}
\mathcal{L} &=& \bar{\nu}_{a} \left(\,i\,\,\slashed{\partial}- m_{a}\,\right) \delta_{ab}\,\nu_{b}
+ \,\bar{\nu}_{a}\,\lambda_a\,\delta_{ab}\,\gamma^0 \, \hat{\nabla} \cdot \vec{\nabla}\,\nu_{b}\,,  \nonumber \\
\eea
where $a,b=1,2,3$.
The corresponding energy dispersion relations for the neutrino mass eigenstates are determined to be
\bea
\label{EnergyNu}
E_a &=&\sqrt{p^2+m_a^2} + \lambda_a \,p \,,~~~~\textrm{for neutrinos,} \\
\label{EnergyAntiNu}
\bar{E}_a &=&\sqrt{p^2+m_a^2} - \lambda_a \, p \,,~~~~\textrm{for anti-neutrinos.}
\eea
We expect $\lambda_a \ll 1$ to
be consistent with current experiments. As a result, neutrino and anti-neutrinos acquire the same mass but slightly
different energy dispersion relations. This is in contrast to the conventional sense of CPT violation in the
neutrino sector, which requires neutrinos and anti-neutrinos to acquire different masses \cite{LSNDCPTV}.

Since neutrino and anti-neutrinos acquire different energy dispersion relations, the usual expansion of field operators
in terms of creation and annihilation operators would have to be modified. The neutrino field operators are defined as
\bea
&&\nu(x) = \int\,\frac{d^3\,\mathbf{p}}{(2\pi)^3}\,\sum_{s}\, \nonumber \\
&& \left(\,a_s(\mathbf{p})\,u_s(p)
\,\frac{e^{-i\,p\,x}}{\sqrt{2\,E_\mathbf{p}}}
+b^{\dagger}_s(\mathbf{p})\,v_s(\bar{p})\,\frac{e^{i\,\bar{p}\,x}}{\sqrt{2\,\bar{E}_\mathbf{p}}}\,\right) \\
&&\bar{\nu}(x) = \int\,\frac{d^3\,\mathbf{p}}{(2\pi)^3}\,\sum_{s}\, \nonumber \\
&& \left(\,b_s(\mathbf{p})\,\bar{v}_s(\bar{p})
\,\frac{e^{-i\,\bar{p}\,x}}{\sqrt{2\,\bar{E}_\mathbf{p}}}
+a^{\dagger}_s(\mathbf{p})\,\bar{u}_s(p)\,\frac{e^{i\,p\,x}}{\sqrt{2\,E_\mathbf{p}}}\,\right)
\eea
where $p^0 = E_{\mathbf{p}}$, $\bar{p}^0 = \bar{E}_{\mathbf{p}}$ and we have suppressed all the flavor or mass indices for generality.
The creation and annihilation operators can be imposed to obey the usual anticommutation relations: $\{a_r(\mathbf{p}),\, a^{\dagger}_s(\mathbf{q})\} = \{b_r(\mathbf{p}),\, b^{\dagger}_s(\mathbf{q})\}
= (2\pi)^3\,\delta_{rs}\,\delta^{(3)}(\mathbf{p}-\mathbf{q}) $.\,
This together with the usual sum rules for the spinors $u_s (\mathbf{p})$ and $v_s(\mathbf{p})$ lead to the \emph{equal-time} anticommutation
relations for the field operators:\, $\{\nu (\mathbf{x}),\, \nu^{\dagger}(\mathbf{y})\} = \delta^{(3)}(\mathbf{x}-\mathbf{y})$\, and \,
$\{\nu(\mathbf{x}),\, \nu(\mathbf{y})\} = \{\nu^{\dagger}(\mathbf{x}),\, \nu^{\dagger}(\mathbf{y})\} = 0$.

If we start from the Lagrangian \eqref{L_flavor} or \eqref{L_mass} and compute the conjugate momentum operator $\pi(x)$
associated with the field operator, we obtain $\pi(x)= i\, \nu^{\dagger}(x)$. The canonical quantization rule
requires \,$\{\nu (\mathbf{x}),\,\pi(\mathbf{y})\} =i\, \delta^{(3)}(\mathbf{x}-\mathbf{y})$\, and hence \,$\{\nu (\mathbf{x}),\, \nu^{\dagger}(\mathbf{y})\} = \delta^{(3)}(\mathbf{x}-\mathbf{y})$.
This is obviously consistent with what we derived from the neutrino field operators $\nu(x)$ and $\nu^{\dagger}(x)$ directly, and so
the internal consistency of the entire construction is established.
Therefore, we conclude that despite CPT violation in our model, spin-statistics is preserved.
In fact, the above discussion reveals that any interaction term that breaks
CPT but does not contain \,$\partial_t\,\nu$\, can preserve spin-statistics. \\


\emph{Leptogenesis.} \,A successful baryogenesis needs a process which satisfies all of the three Sakharov
conditions \cite{Sakharov} simultaneously: \,baryon number violation, C and CP violations, and non-equilibrium condition.
One remarkable way to explain the observed baryon asymmetry in the universe is through leptogenesis.
The main idea is that lepton asymmetry is preferentially generated in the very early universe.
It is then partially transformed into baryon asymmetry by the sphaleron process \cite{Sphalerons} which violates both
baryon number (B) and lepton number (L). The analogous Sakharov conditions for leptogenesis are similar but with
baryon number violation replaced by lepton number violation. In the standard paradigm of leptogenesis \cite{Leptogenesis},
a heavy right-handed Majorana neutrino decays into leptons and Higgs. This decay process is both L-violating and CP-violating.
Interestingly, the right-handed Majorana neutrino is also responsible for explaining the smallness of neutrino masses
through the see-saw mechanism \cite{See-saw}.

In contrast to the standard leptogenesis,
CPT violation allows the Dirac left-handed neutrinos and right-handed anti-neutrinos to develop an asymmetry even at thermal equilibrium:
\bea
\label{n_minus_nbar}
n_{\nu_a}-n_{\bar{\nu}_a} = \int_{0}^{\infty}\,\frac{dp}{2\,\pi^2}\;\,p^2 \left(\, \frac{1}{e^{E_a/T}+1}-\frac{1}{e^{\bar{E}_a/T}+1}\,\right)
 \nonumber \\
\eea
where we have set the Boltzmann constant $k_B=1$ for convenience, $T$ is the temperature, $n_{\nu}$ and $n_{\bar{\nu}}$
are the Fermi-Dirac distribution for neutrinos and anti-neutrinos respectively.
Since spin-statistics is preserved in our model, we are safe to use the Fermi-Dirac distribution.

With a given temperature $T$, the integrand in \eqref{n_minus_nbar} is suppressed unless $p\sim T$. Thus,
if $ \sqrt{\lambda_a}\,T \,\gg \,m_a$ (which will be evidently justified in a moment),
we can approximate \,$E \approx (1+\lambda_a) \,p $\, and \,$\bar{E} \approx (1-\lambda_a) \,p $\, in the integrand.
Performing the integration over
$p$ and keeping only the leading order, we obtain the neutrino asymmetry
\bea
\label{integrate}
n_{\nu_a}-n_{\bar{\nu}_a} \approx -\,\frac{9\,\lambda_a}{2\,\pi^2} \; \zeta (3)\; T^3 \,+\, \mathcal{O}(\,\lambda_a^3\,)\,,
\eea
for $\sqrt{\lambda_a}\,T \,\gg \,m_a$,\, with $\zeta (3) \approx 1.202$ being the Riemann zeta function.

At the thermal equilibrium, the entropy per comoving volume is conserved.
The entropy density is given by \,$s= (2 \pi^2/ 45)\, g_{\ast}\, T^3 $ \cite{Turner}.
For $T \gtrsim$ 100 GeV, we have $g_{\ast} \sim 106$. Thus, the total neutrino asymmetry to entropy density ratio is
\bea
\sum_{a=1}^3\,\frac{n_{\nu_a}-n_{\bar{\nu}_a}}{s} \sim -\,10^{-2} \, \sum^3_{a=1}\,\lambda_a\,.
\eea
A successful leptogenesis requires this ratio to be of order $10^{-10}$, which in turn requires
\bea
\lambda_a \sim 10^{-8}\,.
\eea
This is obviously valid if $\lambda_1=\lambda_2=\lambda_3 =\lambda$ which is implied from the lepton-number preserving case with
a diagonal $\lambda_{\alpha \beta}=\lambda \,\delta_{\alpha \beta}$ in \eqref{L_flavor}.
Thus, even if we keep a general and non-diagonal $\lambda_{\alpha \beta}$ which violates the individual neutrino lepton numbers,
this violation is irrelevant for leptogenesis. We conclude that our model is capable of generating the correct amount of lepton asymmetry
without lepton number violation and
non-equilibrium condition.

As in the standard paradigm of leptogenesis, this pre-existing neutrino asymmetry will be partially converted into baryon
asymmetry through the (B+L)-violating but (B-L)-preserving sphaleron processes (which are significant for  $T \gtrsim 100$ GeV).
When the chemical equilibrium is reached, the baryon asymmetry is equal to \cite{Baryogenesis}
\bea
\label{asymmetry}
\frac{n_{B}-n_{\bar{B}}}{s} = -\,0.35\, \sum_{a=1}^{3}\,\frac{n_{\nu_a}-n_{\bar{\nu}_a}}{s} \sim 10^{-10}\,.
\eea
As the sphaleron processes freeze out below the electroweak scale, this baryon asymmetry will be permanently built into
the quark sector. The quarks are confined to form baryons as the universe cools below the QCD phase transition scale (about 150 MeV),
and this asymmetry becomes what we observe today.

In retrospect, we confirm that since baryogenesis in our model occurs above the electroweak scale, the condition
$\sqrt{\lambda_a}\,T \,\gg \,m_a$ and hence the approximations
\,$E \approx (1+\lambda_a) \,p $\, and \,$\bar{E} \approx (1-\lambda_a) \,p $\,
are justified.

We remark that if, on the contrary, one assumes left-handed neutrinos and right-handed anti-neutrinos to have different masses $m_a$ and $\overline{m}_a$, as was done by
\cite{LSNDCPTV} to resolve LSND, then the baryon asymmetry to entropy density ratio would have gone as
\bea
\frac{n_{B}-n_{\bar{B}}}{s}\sim 10^{-4}\,\sum_{a=1}^{3}\,\frac{m_a^2-\overline{m}_a^2}{T^2}\,.
\eea
For $T \gtrsim 100$ GeV at which sphalerons are effective, we require
$m_a^2-\overline{m}_a^2 \gtrsim $ (100 MeV)$^2$ to ensure a successful baryogenesis,
which is incompatible
with the mass scale suggested by LSND or any other experiments. But we emphasize that our model of CPT-violating neutrinos predicts
the correct amount of baryon asymmetry.

In fact, an earlier idea of CPT-odd leptogenesis has been explored in \cite{Pospelov}. The authors of \cite{Pospelov}
considered non-renormalizable dimension-5 operators (involving heavy Majorana neutrinos) that are CPT-violating and
lepton-number violating. In comparison, our new idea of leptogenesis from CPT violation is unique in the sense that both
new particles and lepton number violation are \emph{not} required.

Furthermore, \cite{Pospelov} listed a set of constraints on CPT-violating dimension-5 operators in the fermionic sector of the
Standard Model \cite{CPTconstraints}. In all of these works, CPT violation is achieved by the existence of a constant background
vector. The constraints on the dimensionful coupling constants are
thus derived. On the contrary, we are considering a renormalizable CPT-violating operator in the current paper. Now,
CPT violation is achieved by the existence of a constant background tensor and the coupling constant is dimensionless.
So it is not obvious that the constraints from \cite{CPTconstraints} are directly applicable to our work.
We plan to find similar constraints in a forthcoming article.  \\

\emph{Implications for Neutrino Experiments.}\, \, In the conventional neutrino oscillation formulae, the oscillation
frequency is proportional to $ \Delta E_{ab} = E_a -E_b$, with $a,b=1,2,3$.
If Lorentz invariance and CPT are both preserved, the conventional energy dispersion holds, and the frequency
oscillation is given by
\bea
\Delta E_{ab} \approx \frac{1}{2 E}\,\Delta m_{ab}^2 \,,
\eea
where $\Delta m_{ab}^2=m_a^2-m_b^2$ and $E \approx  E_a\approx E_b$ because neutrinos are relativistic.
However, in our model, neutrinos and anti-neutrinos acquire the energy dispersions according to
\eqref{EnergyNu} and \eqref{EnergyAntiNu} respectively. This means that
\bea
\Delta E_{ab} &=& \frac{1}{2\,E} \left[\, \Delta m_{ab}^2 + 2\, E^2 \,(\,\lambda_{a}-\lambda_b \,)\,\right] \,,\\
\Delta \bar{E}_{ab} &=& \frac{1}{2\,E} \left[\, \Delta m_{ab}^2 - 2\, E^2 \,(\,\lambda_a-\lambda_b\,)\,\right]\,.
\eea
As a result, to confront our model with experiments, any experimental constraints on
$\Delta m_{ab}^2$ will have to be re-interpreted as constraints on
\bea
\Delta M^2_{ab} (E) &\equiv& \Delta m_{ab}^2 \,+ 2\, E^2 \,(\,\lambda_{a}-\lambda_b \,)\,,\\
\Delta \overline{M}^2_{ab} (E) &\equiv& \Delta m_{ab}^2 \, - 2\, E^2 \,(\,\lambda_{a}-\lambda_b \,)\,.
\eea

Our model only modifies the usual energy dispersions of neutrinos and anti-neutrinos, but \emph{not} the
mixing angles. We adopt the usual mixing angles extracted from solar (SNO), atmospheric (Super-Kamiokande) and
reactor (KamLAND, CHOOZ \cite{Chooz})
neutrino experiments. This means that we take $\sin^2(2\theta_{12})\sim 0.8$, \,$\sin^2(2\theta_{23})\sim 0.9$ \,and \,$\sin^2(2\theta_{13}) < 0.15$.
SNO and KamLAND have measured the survival probabilities of $\nu_e \rightarrow \nu_e$ and $\bar{\nu}_\mu \rightarrow \bar{\nu}_e$
respectively, with both $\nu_e$ and $\bar{\nu}_e$ being in the MeV scale. Besides, Super-Kamiokande (SuperK) has measured the oscillation
probability for atmospheric neutrinos with energy from a few GeV up to 100 GeV (although the detector is not able to distinguish neutrinos from anti-neutrinos
in the flux).

By assuming the usual mixing angles, we are required to satisfy the following
constraints from SNO and KamLAND respectively:
\bea
&& \Delta M^2_{21} (\,\textrm{MeV}\,) \approx 7.6 \times 10^{-5}\, \textrm{eV}^2 \,,\\
&& \Delta \overline{M}^2_{21} (\,\textrm{MeV}\,) \approx 7.6 \times 10^{-5}\, \textrm{eV}^2 \,.
\eea

Apparently, MINOS may be explained in the following way. In MINOS, the average neutrino or anti-neutrino energy is about GeV.
So the constraints on the mass-squared splittings are translated into
\bea
\label{MINOS1}
 |\,\Delta M^2_{32} (\,\textrm{GeV}\,)\,| &=&  2.35\times 10^{-3}\, \textrm{eV}^2 \,,\\
\label{MINOS2}
|\,\Delta \overline{M}^2_{32} (\,\textrm{GeV}\,)\,| &=& 3.36\times 10^{-3}\, \textrm{eV}^2 \,.
\eea
For normal mass hierarchy ($\Delta m_{32}^2 >0$), the above
conditions \eqref{MINOS1} and \eqref{MINOS2} can be satisfied if $\Delta m_{32}^2 \approx 2.86 \times 10^{-3}\, \textrm{eV}^2 $ and $\lambda_2- \lambda_3 \approx 2.5 \times 10^{-22}$.
For inverted mass hierarchy ($\Delta m_{32}^2 <0$), these
conditions are satisfied if $ \Delta m_{32}^2  \approx - 2.86 \times 10^{-3}\, \textrm{eV}^2 $ and $\lambda_3- \lambda_2 \approx 2.5 \times 10^{-22}$.
The validity of using the 2-neutrino oscillation formula to perform the data fitting also requires
$\Delta M^2_{21} (\,\textrm{GeV}\,)\approx \Delta \overline{M}^2_{21} (\,\textrm{GeV}\,) \approx 7.6 \times 10^{-5}\, \textrm{eV}^2$\,.
Thus, MINOS can be explained without upsetting the solar and reactor neutrino data if
\bea
\label{MINOS_deltam2}
&& \Delta m_{21}^2 \approx 7.6\times 10^{-5}\, \textrm{eV}^2 ~~;~~ |\,\Delta m_{32}^2\,| \approx 2.86 \times 10^{-3}\, \textrm{eV}^2  \nonumber \\
&& \lambda_1=\lambda_2 ~~;~~ |\lambda_3- \lambda_2| \approx 2.5 \times 10^{-22}\,.
\eea

However, the parameters in \eqref{MINOS_deltam2} would imply
\bea
 |\,\Delta M^2_{32} (\,100\,\textrm{GeV}\,)\,|\sim |\,\Delta \overline{M}^2_{32} (\,100\,\textrm{GeV}\,)\,| \sim \,
 \textrm{eV}^2\,.
\eea
Since the SuperK data indicate a rather flat oscillation spectrum up to the high energy region ($\sim$100 GeV), the parameters
in \eqref{MINOS_deltam2} obviously predict too many oscillations for high energy neutrinos or anti-neutrinos at SuperK. Therefore,
they are excluded by the SuperK data.

The only way for our model to be consistent with \emph{all} of the solar, atmospheric and reactor neutrino data is the simplest
case with $\lambda_1=\lambda_2=\lambda_3=\lambda$. As mentioned earlier, this corresponds to the lepton-number preserving case with
a diagonal $\lambda_{\alpha \beta}=\lambda \,\delta_{\alpha \beta}$ in \eqref{L_flavor}.
In this case, leptogenesis is still explained, although we will have
\bea
\Delta M^2_{ab} (E) = \Delta m^2_{ab} = \Delta \overline{M}^2_{ab} (E) = \Delta \overline{m}^2_{ab}\,.
\eea
So the effect of CPT violation is completely invisible in all the neutrino oscillation experiments.
The frequencies of neutrino oscillations predicted by our model and the conventional
theory of neutrino oscillation are exactly the same. Of course, this would imply that our model cannot explain MINOS and
the $\bar{\nu}_\mu \rightarrow \bar{\nu}_e$
``anomaly" in LSND and MiniBooNE. \\


\emph{Conclusion.}\,
We construct a new model in which neutrinos and anti-neutrinos
acquire the same mass but slightly different energy dispersion relations.
This simple model of neutrino CPT violation explains leptogenesis easily, without
lepton number violation and the non-equilibrium condition.
Also, it is consistent with all of the solar, atmospheric and reactor neutrino data.
In addition, according to FIGURE 13.10 in \cite{PDG}, our model is also consistent with all other neutrino experiments such as
KARMEN (40 MeV), Bugey (MeV), CDHSW (GeV), NOMAD (50 GeV), Palo Verde (MeV), etc.

It would be interesting to generalize the idea of the current model and see if MINOS and the
$\bar{\nu}_\mu \rightarrow \bar{\nu}_e$ ``anomaly" in both of LSND and MiniBooNE can be explained as
well. We will explore this possibility in a forthcoming article.
\\



\emph{Acknowledgments.}\,
We sincerely thank Alan Kostelecky, Tom Weiler, Pierre Ramond and Joachim Kopp for useful discussions. We also
thank an anonymous referee for useful comments.
This work was supported by
US DOE grant DE-FG05-85ER40226.

\end{document}